\documentstyle[12pt]{article}

\begin{document}
\def\a{{\bar{\alpha}}_s}
\def\be{\begin{equation}}
\def\ee{\end{equation}}
\hoffset = -1truecm\voffset = -2truecm

{}
\vskip 1cm

\hfill{KEK-preprint 96-143} 

\hfill{TH-499}

\vskip 1cm
\begin{center}
{\Large \bf
Theoretical uncertainties for weak decays: 

higher dimension operators}

\vskip 1cm

{\bf A.A.Pivovarov}

{\it Institute for Nuclear  Research of the Russian Academy of Sciences

Moscow 117312, Russia}\footnote{Permanent address}

{\rm and}

{\it National Laboratory for High Energy Physics (KEK)}

{\it Tsukuba, Ibaraki 305, Japan}

\vskip 1cm
{\bf Abstract}
\end{center}
A brief review of recent results on computing 
contributions of higher dimension operators to weak 
effective $\Delta S=1,2$ hamiltonians for light quarks
is presented.

\vspace{0.5in}

\vskip 3.3cm
\begin{center}
Talk given at Fourth KEK Topical Conference on Flavor Physics,

KEK, Tsukuba, 305 Japan, 29-31 October, 1996
\end{center}\thispagestyle{empty}
\newpage
\noindent {\bf 1. Introduction}

I describe some of our recent results on computing
contributions of higher dimension 
operators to effective $\Delta S=1,2$ hamiltonians within the Standard Model.
These corrections are due to higher order terms in heavy quark mass expansion 
($m_c^{-1}$) and require thorough numerical study before being 
safely neglected.
\vskip 0.3cm
\noindent {\bf 2. Corrections to $K\rightarrow \pi\pi$ decays: 
$\Delta S=1$ effective hamiltonian.} 

The short-distance analysis after removing
the $W$-boson and the heavy quarks results in the 
effective $\Delta S=1$ hamiltonian of the
following  form \cite{effham0,effham1}
\be
H_{\Delta S=1}= {G_F\over\sqrt
2}V_{ud}V^*_{us}\sum^{6}_{i=1}
[z_i(\mu)+\tau y_i(\mu)Q_i]+h.c.
\label{effham0}
\ee
where $G_F$ is the Fermi constant, $V$ stands for
the Cabibbo-Kobayashi-Maskawa mixing matrix,
$\tau=-{V_{td}V^*_{ts}/V_{ud}V^*_{us}}$,
$z_i(\mu)$ and $y_i(\mu)$ are Wilson coefficients,
and $\{Q_i|i=1,...,6\}$ is the basis of local
operators 
of dimension six in mass units (four-quark
operators). In Eq.~(\ref{effham0})
contributions of electroweak penguin operators 
\cite{empeng0,empeng1} are omitted.
For more precise comparison of theoretical predictions with experiment
the existing analysis
cannot be considered complete and
following points require further investigation:
\begin{itemize}
\item Perturbation theory for
kaon nonleptonic decays can be improved  by
using more accurate effective hamiltonian. Within the
standard approach only leading terms in the
inverse masses of heavy particles are kept
while a proper account of
nonleading corrections in the inverse mass  of
charmed quark \cite{p1}	 
can be
important.  
Further corrections to the effective hamiltonian
appear because the top quark is heavier than other
quarks and $W$-boson. This results in an
incomplete GIM cancellation
and in appearance of a
new operator in the effective hamiltonian \cite{heavytop}.
\item
High precision
theoretical estimates for kaon decays stumble
at a necessity to calculate mesonic matrix
elements of local four-quark operators entering the
effective hamiltonian (\ref{effham0}). The only method of
computation  entirely based on first principles
seems to be  numerical simulations on the lattice
(see, {\it e.g.} \cite{mart}) though so far even this approach has not given
unambiguous and stable results.
Several
semi-phenomenological techniques have been
developed and applied for computation of
matrix elements, {\it e.g.} \cite{1nc}, though
the precision still need to be essentially
improved.
Recently a regular method to evaluate the
mesonic matrix elements has been exploited
\cite{p2} where the effective hamiltonian is
represented in terms of the chiral
theory variables \cite{gasser}
and parameters of the chiral
representation are determined via QCD sum
rules \cite{svz,kp,kpt} for an appropriate three-point Green's
function. 
\item
Perturbation theory does not take into account
soft light quarks and
gluons with small virtual momenta. They
are entirely
hidden in matrix elements of
local four-quark operators. Well known
factorization procedure for evaluation of these
matrix elements \cite{fact} accounts only for the
"factorizable" part of the interaction \cite{bb3,bb2,kkal}.
"Unfactorizable" contributions, for example, those
corresponding to annihilation of a quark pair
from the four-quark operator into soft gluons \cite{NuoCim}
are omitted.
The calculation of these contributions
and the generalization of the matrix element estimates
beyond the factorization framework can be
systematically done within the approach of ref.~\cite{koko}.
\end{itemize}
\noindent
After decoupling heavy particles
($W$-boson, $t$-, $b$-, and $c$-quarks) from the
light sector of the theory, Eq.~(\ref{effham0})
corresponds to the leading order in inverse masses
of these particles. The removal of the $c$-quark
however is not very reliable and, in general,
requires a special investigation. It is not heavy
enough in comparison with a characteristic mass
scale  in  the sector of light $u$-, $d$-, and
$s$-quarks, for example, with the $\rho$-meson
mass.  The nonleading terms in the $1/m_c$
expansion can, therefore, be important  and
require a quantitative consideration.
To compute corrections of order $1/m_c$,
the tree level hamiltonian
before decoupling
of the $c$-quark is used. It reads
\be
H^{tr}_{\Delta S=1}= {G_F\over\sqrt
2}V_{ud}V^*_{us}(Q_2^u-(1-\tau )Q_2^c)+h.c.
\label{effhamtree}
\ee
where $Q_2^q=4(\bar s_L\gamma_\mu q_L)(\bar
q_L\gamma_\mu d_L)$, $q_{L(R)}$ stands for
left(right) handed quark.  Performing the
OPE and restricting oneself to the first order
terms in $\alpha_s$ and $m_c^{-2}$ one finds
\be
H_{\Delta S=1}=H^{(6)}+H^{(8)}.
\label{split}
\ee
The  first addendum in Eq.~(\ref{split}) $H^{(6)}$ corresponds to
leading contributions in $1/m_c$ and coincides
with Eq.~(\ref{effham0}). Second addendum
in Eq.~(\ref{split}) is the
$1/m_c$ correction \cite{p1}
\be
H^{(8)}= {G_F\over\sqrt 2}V_{u
d}V^*_{us}(1-\tau){\alpha_s\over 4\pi m_c^2}
\left(\sum^7_{i=1} C_i^{(8)} Q_i^{(8)}+
\sum^4_{i=1} C_i^{(7)} m_sQ_i^{(7)}\right) +h.c.
\label{effham1}
\ee
where a basis $\{ Q^{(8)}_i|i=1,...,7\}~ (\{
Q^{(7)}_i|i=1,...,4\})$ of  the  local  operators
with dimension eight (seven) in mass units is
chosen in  the form
$$
Q_1^{(8)} =\bar s_L(\hat DG_{\mu\alpha}G_{\nu\mu}
\sigma_{\alpha\nu}+G_{\nu\mu}\sigma_{\alpha\nu}
\hat DG_{\mu\alpha})d_L,
$$
$$
Q_2^{(8)} =ig_s\bar s_L(J_\mu\gamma_\alpha
G_{\alpha\mu}- \gamma_\alpha
G_{\alpha\mu}J_\mu)d_L,
$$
$$
Q_3^{(8)}
=\bar s_L(P_{\alpha}G_{\mu\alpha}\gamma_\nu
G_{\nu\mu}+\gamma_\nu G_{\nu\mu}
G_{\mu\alpha}P_{\alpha})d_L,
$$
$$
Q_4^{(8)} =g_s\bar
s_L(G_{\mu\nu}\sigma_{\mu\nu}\hat J+ \hat J
G_{\mu\nu}\sigma_{\mu\nu})d_L,
$$
$$
Q_5^{(8)}
=i\bar s_L(G_{\mu\nu}\sigma_{\mu\nu}\gamma_ \alpha
G_{\alpha\beta}P_\beta -P_\beta\gamma_\alpha
G_{\alpha\beta} G_{\mu\nu}\sigma_{\mu\nu})d_L,
$$
$$
Q_6^{(8)} =\bar s_L(D^2\hat J)d_L, ~~Q_7^{(8)}
=i\bar s_L(\hat DG_{\nu\mu}G_{\nu\mu}-
G_{\nu\mu}\hat DG_{\nu\mu})d_L,
$$
$$
Q_1^{(7)}
=\bar s_R(G_{\mu\nu}\sigma_{\mu\nu}
G_{\alpha\beta} \sigma_{\alpha\beta})d_L,~~
Q_2^{(7)}=\bar s_R(G_{\mu\nu}G_{\nu\mu})d_L,
$$
\be
Q_3^{(7)} =i\bar s_R(G_{\nu\alpha}G_{\alpha\mu}
\sigma_{\nu\mu})d_L,~~ Q_4^{(7)}=\bar s_R(J_\mu
P_\mu +P_\mu J_\mu)d_L.
\label{basis}
\ee
Here $P_\mu  =i\partial_\mu+g_sA_\mu$ is the
momentum  operator  in  the  presence  of
external field $A_\mu \equiv A_\mu^a t^a$,
$t^a$ are generators of the color
group $SU(3)$, $G_{\mu\nu}\equiv G^a_{\mu\nu}t^a$
is  the gluon field strength tensor, $J_\mu\equiv
\sum_{q=u,d,s} (\bar  q\gamma_\mu t^a q)t^a$, and
$\sigma_{\mu\nu}=i[\gamma_\mu,\gamma_\nu]/2$.
In derivation of Eq.~(\ref{effham1}) $u$- and
$d$-quark are considered massless, and the first order
in strange quark mass was kept with the use of
equations of motion
$
\bar  s\hat  P=m_s\bar s,~~~\hat P d=0,\quad
[P_\mu, G_{\mu\nu}]=iD_\mu
G_{\mu\nu}=-ig_sJ_\nu .
$
Straightforward calculation gives
the following values for coefficients
$C_i^{(j)}$ to the leading order in $\alpha_s$
\[
C_2^{(8)}=-2 C_1^{(8)}
=4C_3^{(8)}=-8C_4^{(8)}=2 C_6^{(8)}=8C_7^{(8)}=-{16\over 15},\quad
C_5^{(8)}=0,
\]
\be
C_1^{(7)}=C_2^{(7)}=-{2\over 5},
~~C_3^{(7)}={6\over 5},~~C_4^{(7)}=0.
\label{coef}
\ee
This generalizes the effective hamiltonian
for $\Delta S=1$ decays beyond the leading order
in  $1/m_c$ expansion.

To complete our treatment of the local effective
hamiltonian we consider the case of a heavy top quark.
GIM cancellation is not perfect
and the quark-gluon operator $m_sQ^{(5)}$ appears in
the effective hamiltonian already in the first
order in $\alpha_s$. This additional contribution
\cite{heavytop} reads
$$
\Delta H^{(6)}={G_F\over\sqrt
2}V_{ud}V^*_{us}\tau
C^{(5)}(\mu )m_sQ^{(5)}(\mu ),
$$
$$
C^{(5)}(\mu )={1\over 16\pi^2}
\left(F(x_c)-F(x_t)\right)\eta (\mu ),\quad x_q={m_q^2\over M_W^2},
$$
$$
F({x_q})={1\over 3}{1\over (x_q-1)^4}
\left({5\over 2}{x_q}^4-7{x_q}^3+
{39\over 2}{x_q}^2-19{x_q}+4-
9{x_q}^2\ln{x_q}\right),
$$
\be
F(x_c)\sim F(0)={4\over 3}.
\label{heavytop}\ee
The renormalization group factor has the form
$$
\eta(\mu )
=\left({{{\bar{\alpha}}_s}(m_b)
\over{{\bar{\alpha}}_s}(M_W)}\right)
^{\gamma^{(5)} /{2\beta_5}}
\left({{{\bar{\alpha}}_s}(m_c)
\over{{\bar{\alpha}}_s}(m_b)}\right)
^{\gamma^{(5)} /{2\beta_4}}
\left({{{\bar{\alpha}}_s}(\mu
)\over{{\bar{\alpha}}_s}(m_c)}\right)
^{\gamma^{(5)} /{2\beta_3}}
$$
where $\gamma^{(5)}=-28/3$ is the anomalous
dimension of the operator $m_sQ^{(5)}$ \cite{mor},
$\beta_{n_f}=11-{2\over3}n_f$, $n_f$ is the number
of active quarks flavors.

Contributions of $u$- and $c$-quarks to
the real part of the effective hamiltonian cancel
each other via GIM mechanism and the operator
$m_sQ^{(5)}$ contributes to imaginary
parts of amplitudes and, therefore,
can be important in the analysis of direct
CP violation.  However, its contribution is
suppressed numerically because $\eta(\mu )<1$ and
the function $F(x)$ changes slowly.  Indeed, at
the point $\Lambda_{\rm QCD}=0.3$~GeV, $\mu=1$~GeV,
$m_t=130$~GeV, one has $C^{(5)}=0.001$, while the
numerical value of the Wilson coefficient of
the dominant penguin operator
$Q_6=-8\sum_{q=u,d,s}(\bar s_L q_R)(\bar q_R d_L)$
is $y_6=0.1$.

Thus, the complete form of the effective
hamiltonian up to the first order in $m_s$,
$1/m_c$ and $\alpha_s$ with $m_t\sim M_W$ is
now available. 

As an example of using the above hamiltonian we
consider $K\rightarrow\pi\pi$ decays. For
this end we have  to  extract  an  information
about the matrix elements  of  the  local
operators $Q^{(j)}$ between mesonic states.
We start with operators $m_sQ^{(7)}_i$ and
$m_sQ^{(5)}$.  These operators contain explicitly
the strange quark mass and, therefore, in the
leading order of the chiral expansion they
correspond to tadpole terms in the chiral weak
lagrangian \cite{weakch} that
do not generate any observable effect
and can be neglected in the leading order of
chiral symmetry  breaking \cite{crew,desh}.

Thus the problem is reduced to the estimation of
the matrix elements  of the operators $Q^{(8)}_i$.
At present there is no regular method to calculate
them within QCD except
direct simulations on the lattice. To estimate
at least the scale of nonleading $1/m_c$
corrections we work with a simplified
model that uses factorization.
One selects operators
containing scalar quark currents which can be
written as
$
(\bar s_L
G_{\mu\nu}\sigma_{\mu\nu}q_R)(\bar q_R d_L)
$ and
$(\bar s_L q_R)(\bar q_R
G_{\mu\nu}\sigma_{\mu\nu}d_L).
$
This step  seems to  be justified because in  the
case  of  dimension  six operators  the   similar
"penguin-like" structures are strongly enhanced
and dominate the others.  The   last
simplification   consists in the substitution
$
\bar q g_s G_{\mu\nu}\sigma_{\mu\nu}q
\rightarrow m_0^2\bar q q
$
where $m_0^2$ determines the scale of nonlocality
of the quark condensate and is defined by the
equation $\langle \bar q g_s G_{\mu\nu}\sigma_
{\mu\nu}q\rangle =m_0^2\langle \bar qq\rangle ,
\quad m_0^2(1~{\rm GeV})=0.8\pm 0.2~{\rm GeV}^2$ \cite{m02,m02op}.
This substitution is valid in the chiral limit for the
operator $\bar q g_s G_{\mu\nu}\sigma_{\mu\nu}q$. We suppose that
it  is justified also in our case at least for
estimates of the order of magnitude.
All above assumptions about the factorization
procedure in the case of dimension six operators
become exact within the many color limit of
QCD, $N_c\rightarrow\infty$, to the leading order
in $N_c$ \cite{thooft}.

Thus, only operator $Q^{(8)}_4$ has a nonvanishing matrix
element that reads
\be
\langle \pi\pi|Q_4^{(8)}|K\rangle ={m^2_0\over 4}
\langle \pi\pi|Q_6|K\rangle .
\label{me}
\ee
We should note that the coefficients  $C_i^{(8)}$
are finite to the leading order in $\alpha_s$
and independent of renormalization scheme. We can
therefore use the leading order values of
mesonic matrix elements that is consistent up to the
considered level of accuracy. The
next-to-leading $\alpha_s$ corrections to Wilson
coefficients depend on the renormalization scheme
and matching between them 
and mesonic matrix elements is necessary to make physical amplitudes 
independent of
the renormalization scheme.
Thus, account for the first  order
$1/m_c$ corrections results in the
shift of coefficients in front of  the
penguin operator $Q_6$
\be
z_6\rightarrow
\left(z_6+{\alpha_s\over 4\pi}{m^2_0 \over
4m^2_c}C^{(8)}_4\right),
~~~~~y_6\rightarrow
\left(y_6-{\alpha_s\over 4\pi}{m^2_0 \over
4m^2_c}C^{(8)}_4\right).
\label{res1}
\ee
Using   the    numerical   values    $z_6=-0.015$,
$y_6=-0.102$ at the point
$\Lambda_{\rm QCD}=0.3$~GeV, $\mu=1$~GeV,
$m_t=130$~GeV one finds numerically
relative corrections to the Wilson
coefficients to be 
$$
z_6\rightarrow z_6(1-0.1),\quad y_6\rightarrow y_6(1+0.01).
$$
The main correction appears in the real part  of
the  Wilson  coefficient  of the penguin operator
$Q_6$.  Parametrically, the contribution of
dimension eight operators can be as large as 50\%
($m^2_0/m^2_c \sim 0.5$) of the leading term.
However 
much smaller value
has been found within the  simplest
factorization framework for 
meson matrix elements of nonleading $1/m_c$
contributions to the kaon decay amplitudes.
Large violation of the factorization for matrix
elements of dimension eight operators seems
to be likely and the numerical value of
nonleading $1/m_c$ correction can be estimated
only when a self-consistent method to
calculate these matrix elements within QCD
will be available.
\vskip 0.3cm
\noindent{\bf 3. Corrections
to $K^0-\bar{K^0}$ mixing: $\Delta S=2$ effective lagrangian.}

The effective local $\Delta S=2$
lagrangian for the $K^0-\bar K^0$ 
mixing in the leading approximation in the
heavy charmed quark mass $m_c$ is well known \cite{ds21,ds22}.
The corresponding hadronic matrix element
of the effective local $L_{\Delta S=2}$
lagrangian between $K^0$ and $\bar K^0$ states
$
\langle\bar K^0(k')|L_{\Delta S=2}|K^0(k)\rangle
$
has been intensively studied during several last years with different
techniques ({\it e.g.} \cite{kkal,koko,jdon,pichraf,bisan,ld}).
However it
has been pointed out in \cite{wolf}
that the local effective hamiltonian
does not exhaust the physics of $\Delta S=2$ transitions. It cannot
account for the long distance contribution which is present in the initial
Green's function for the matrix element of the $K^0-\bar K^0$ mixing and is
connected with the propagation of the light $u$-quark round the loop of the
box diagram. This contribution is purely nonperturbative and ultimately
depends on infrared properties of QCD. 

Nevertheless there is one point which can be essentially improved
just within perturbation theory for the standard model. It consists in the
calculation of corrections in the inverse mass of charmed quark to the
local part of the effective lagrangian \cite{kkjetppl}. 
These corrections are represented by
local operators with dimension eight in mass units.

Because the $\Delta S=1$ lagrangian
has the form
\be
L_{\Delta S=1}={G_F\over \sqrt2}J_\mu J_\mu^{+}, 
\label{ds2}
\ee
where $J_\mu=\bar{Q_L}\gamma_{\mu}Vq_L$ is
the weak charged hadronic current, $Q=(u,c,t)^T$, $q=(d,s,b)^T$,
the matrix element $M$ of the transition is
represented by
$$
_{out}\langle\bar K^0(k')|K^0(k)\rangle_{in}=i(2\pi)^4\delta (k-k')M,
$$
\be
M={i\over2}\int dx\langle\bar K^0(k')|TL_{\Delta S=1}(x)
L_{\Delta S=1}(0)|K^0(k)
\rangle. 
\label{kkme}
\ee
Eqs. (\ref{ds2}-\ref{kkme}) are valid for the $t$-quark much lighter 
than the $W$-boson. This is not the case anymore but 
for our purpose it is
inessential and in the following we neglect the $t$-quark admixture
and restrict ourselves to the simplified model with two generations.

The effective  $\Delta S=2$  lagrangian can be written in the form
\be
L_{\Delta S=2}
=\left({4G_F \sin\theta_c \cos\theta_c\over\sqrt{2}}\right)^2(L_H+L_L)
\label{nonloc}
\ee
where $\theta_c$ is the Cabibbo angle.
Here
$$
L_H=i\int T_H(x)dx,~~T_H=T_{cc}-T_{cu}-T_{uc}
$$
is the heavy part of the whole effective  $\Delta S=2$  lagrangian containing
loops with virtual heavy $c$-quark in the intermediate state, while
$$
L_L=i\int T_L(x)dx,~~T_L=T_{uu}
$$
describes the light part of the transition.
We introduced useful notations
\[
T_{cu}(x)=T\bar{s}_L \gamma_{\alpha}u_L \bar{c}_L\gamma_{\alpha}d_L(x)
\bar{s}_L \gamma_{\beta}c_L \bar{u}_L\gamma_{\beta}d_L(0)
\]
and so on.
Both $L_H$ and $L_L$
separately require some regularization because they are ultraviolet divergent.
The dimensional regularization is not convenient in this case due to 
the presence
of the $\gamma_5$-matrix. The Pauli-Willars regularization introduces a
regulator mass that makes difficult to perform explicit
calculations. We will use the
regularization which is free of these shortcomings in our
particular case. Namely, let me define the regularized quantities $L^R_{H,L}$
by the equation
$$
L^R_{H,L}=i\int T_{H,L}(x){(-\mu^2x^2)}^{\epsilon}dx
$$
where $\epsilon$ is a regularization parameter and $\mu$ represents the mass
scale analogous to one of dimensional regularization.

Now for the heavy part of the effective $\Delta S=2$ lagrangian $L_H$ 
we develop a
regular expansion in the inverse charmed quark mass in the following form (from
now on we omit the index "$R$")
$$
16\pi^2L_H=C_0(\mu,m_c) O_0(\mu)+\sum_{j}C_j(\mu,m_c)O_j(\mu)
$$
where $C_j$ are coefficient functions depending  on the heavy quark mass
$m_c$ and $O_j$ are  the local operators  built from the light $(u,d,s)$ quark
fields only. If we split the whole lagrangian
into the sum
\be
L_H=L_H^{(0)}+L_H^{(1)}
\label{splitkk}
\ee
then
$
16\pi^2L_H^{(0)}=-m_c^2(\bar s_L \gamma_\alpha d_L)^2 
$
is the well known result of Gaillard and Lee \cite{fact}. 
The rest part of Eq.~(\ref{splitkk})
contains local operators which have dimension eight in mass units.
A convenient form of the operator basis is
\[
O_{\tilde F} = \bar s_L \gamma_\alpha d_L \bar s_L
\gamma_\mu\tilde F_{\mu\alpha} d_L, \quad
O_A=\bar s_L \gamma_{(\mu}D_{\nu)}d_L \bar s_L
\gamma_{(\mu} D_{\nu)}d_L,
\]
$$
O_B=\bar s_L
\gamma_\mu D_\mu d_L \bar s_L \gamma_\nu D_\nu d_L,
$$
\[
O_C= \bar s_L \gamma_\alpha d_L
\bar s_L(\gamma_\mu D_\mu D_\alpha+D_\alpha\gamma_\mu D_\mu)d_L -
{(m_s ^2+m_d^2)\over2} (\bar s_L \gamma_\alpha d_L)^2
\]
where
$
\gamma_{(\mu}D_{\nu)}=(\gamma_{\mu}D_{\nu} +\gamma_{\nu}D_{\mu})/2.
$
The direct calculation gives
$$
16\pi^2L_H^{(1)}=-{4\over3}(O_{\tilde F}+O_A)
\left({1\over\epsilon}+\ln\left({4{\mu}^2e^{-2C}\over {m_c^2}}\right)
+{4\over3}\right)
-{2\over3}O_A
$$
\be
-{2\over3}(O_B+O_C)
\left({1\over\epsilon}
+\ln \left({4\mu^2e^{-2C}\over m_c^2}\right)+{11\over6}\right)
\label{kkmain}
\ee
where $C=0.577...$ is the Euler constant.
After performing
a renormalization (say, a minimal subtraction of the pole term) we will have
the finite quantity and the parameter 
$\mu$ recalls the necessity 
to have the proper short distance contribution 
of $u$-quark.
In this order of the expansion in $m_c^{-1}$ the dependence on this
parameter is explicit contrary to the leading order that is finite
and does not depend on $\mu$. 
The
heavy and light parts of the whole lagrangian must be defined simultaneously in
the coordinated way. The pole part of Eq.~(\ref{kkmain}) is cancelled by the
corresponding divergences of the light part due to GIM mechanism. 
Consider now the light part. The operator product expansion for the
amplitude $T_L(x)$ in $x^2$ at $x^2\rightarrow 0$ has the form
\be
T_L(x)
={1\over 4\pi^4x^6} (\bar{s}_L \gamma_{\alpha}d_L)^2 
+{1\over 24\pi^4 x^4} \left(2O_{\tilde F}+
2O_A+O_B+O_C\right). 
\label{lope}
\ee
The short distance contribution of the light part does cancel 
divergences of the heavy part. More technically we extract the short distance
contribution of the light part by splitting the entire light part into
a sum
\be
L_L =i\int T_L(x){(-\mu^2x^2)}^{\epsilon}dx
=i\int T_L(x)(f(x,\bar x)+{\bar f}(x,\bar x)){(-\mu^2x^2)}^{\epsilon}dx=
L_L^{SH}+L_L^{LG} 
\label{lsplit}
\ee
where
\[
L_L^{SH} =i\int T_L(x)f(x,\bar x){(-\mu^2x^2)}^{\epsilon}dx,  \quad
L_L^{LG}=i\int T_L(x){\bar f}(x,\bar x){(-\mu^2x^2)}^{\epsilon}dx,
\]
and
$
f(x,\bar x)+{\bar f}(x,\bar x)=1.
$
The smooth generalization of the step
$\theta$-function
is chosen for functions $f(x,\bar x)$ and
${\bar f}(x,\bar x)$
$$
f(x,\bar x)={\bar x^{2n}\over \bar x^{2n}+(-x^2)^n},~~
\bar f(x,\bar x)={(-x^2)^n\over \bar x^{2n}+(-x^2)^n}
$$
in such a way that the function $f(x,\bar x)$ cuts out the short
distances only (up to $\bar x$) and the function $\bar f(x,\bar x)$
does the long distances.
The short distance contribution of
the light part is
\[ 
16\pi^2L_L^{SH} = (\bar s_L \gamma_\alpha d_L)^2
{\pi/n\over \sin(\pi/n)} {4\over\bar x^2} 
+{2\over3}\left(2O_{\tilde F}+2 O_A + O_B+O_C\right) 
({1\over\epsilon}+\ln \mu^2\bar x^2).
\]
Now the whole answer is
$$
16\pi^2(L_H+L_L)= 16\pi^2L_L^{LG}+
(\bar s_L\gamma_\alpha d_L)^2\left(-m_c^2 
+{\pi/n\over \sin(\pi/n)}{4\over\bar x^2}\right)
$$
\be
-{2\over3}\left(2O_{\tilde F}+2 O_A+O_B+O_C\right)
\left(\ln({4e^{-2C}\over m_c^2\bar x^2})+{4\over3}\right)-{2\over3}O_A.
\label{finansw}
\ee

The long distance part of the whole light lagrangian $L_L^{LG}$
cannot be calculated due to strong infrared problems and
requires some low-energy model, lattice 
or chiral effective theory \cite{georgi}.
Numerical estimates for corrections 
depend on kaon-antikaon matrix elements of the local
operators $O_{\tilde F} - O_C$ for which 
factorization was used
$$
\langle\bar K^0(k)|(\bar{s}_L \gamma_\alpha d_L)^2|K^0(k)\rangle^{fact}=
(1+{1\over N_c})\left({f_K^2m_K^2\over2}\right)\equiv Z,
$$
\[
\langle\bar K^0(k)|O_{\tilde F}|K^0(k)\rangle^{fact}
=-\delta^2 \left({{f_K^2m_K^2}\over2}\right),
\quad \langle\bar K^0(k')|O_A|K^0(k)\rangle^{fact}=
-\delta^2{1\over N_c}\left({f_K^2m_K^2\over2}\right)
$$
where the parameter $\delta^2$ is defined by the relation \cite{nov}
$
\langle 0|\bar{s}_L \gamma_{\mu}{\tilde F}_{\mu\alpha}d_L|K^0(k)\rangle
=-ik_\mu f_K \delta^2
$
and all other matrix elements vanish.
From Eq.~(\ref{finansw}) one finds
$$
16\pi^2(L_H+L_L^{SH})=
%\langle\bar K^0(k)|(\bar{s}_L \gamma_{\alpha}d_L)^2|K^0(k)\rangle^{fact}
Z(-m_c^2 +{\pi/n\over \sin(\pi/n)}{4\over\bar x^2}
+{4\over3}\delta^2 (\ln({4e^{-2C}\over m_c^2\bar x^2})
+{4\over3})+{1\over6}\delta^2).
$$
Let me estimate $\bar x^2$. Eq.~(\ref{lope}) gives
\be
T_L(x)= {1\over{4\pi^4x^6}}
%\langle\bar K^0(k)|(\bar{s}_L \gamma_{\alpha}d_L)^2|K^0(k)\rangle^{fact}
Z(1-{{\delta^2{x^2}}\over3}+o(x^2)). 
\label{xbar}
\ee
Then at $\delta^2\bar x^2=3$ where the expansion (\ref{xbar}) blows up
and at $n=\infty$ we obtain the following representation for
nonleading corrections
$$
-m_c^2
%\langle\bar K^0(k)|(\bar s_L \gamma_\alpha d_L)^2|K^0(k)\rangle^{fact}
Z\left(1-{4\over3}{\delta^2\over m_c^2}
-{4\over3}{\delta^2\over m_c^2} 
(\ln({4\delta^2\over 3 m_c^2})-2C+{35\over24})\right).
$$
Numerically, at $\delta^2=m_0^2/4$ \cite{m02op} we get
$$
-m_c^2Z(1-0.2+0.3)=-m_c^2Z(1+0.1).
$$
Defining the value of $\bar x^2$ by the relation $\delta^2\bar x^2=1$
we get
$$
-m_c^2Z(1-0.5+0.1)=-m_c^2Z(1-0.4).
$$

\vskip 0.3cm
\noindent{\bf 4. Conclusions.}

The corrections in the inverse mass of the
charmed quark to the effective $\Delta S=1,2$
lagrangians can be important numerically for precious
comparison theory with experiment.
The $\Delta S=2$
lagrangian reveals at this level 
an explicit dependence on the
boarder between short and long distances. 
The OPE at $x^2\rightarrow 0$
for the light part of the lagrangian is analyzed for determining 
the convergence scale.
Within vacuum dominance approximation this scale is
large enough and is given
by the parameter $\delta$. 
The corrections
can change the transition matrix element 
considerably
depending on the concrete choice of the scale $\bar x$.
This means that the
long distance hadronic contribution is important to stabilize the
result in the region where the expansion (\ref{xbar}) 
fails, in other words
the integral $L_L^{LG}$ changes quickly for $\bar x^2$
in the region 
$1/\delta^2<\bar x^2< 3/\delta^2$ in order to compensate
corresponding changes of the short distance part (\ref{lope}). 

\vskip 0.3cm
\noindent
{\bf Acknowledgments.}

\noindent
I thank organizers of the Conference 
for giving me an opportunity to deliver this talk.
I am indebted to Prof.~Y.Okada 
and to colleagues from Theory Group
for the great hospitality extended to me during
my stay at KEK where the final version of the paper was written.  
This work is supported in part
by Russian Fund for Basic Research No. 96-01-01860.

\end{document}